\documentclass[
 twocolumn,
 reprint,
 superscriptaddress,
 amsmath,amssymb,
 aps,
 prd,
]{revtex4-2}

\usepackage{soul}
\usepackage[normalem]{ulem}
\usepackage{orcidlink}
\usepackage{graphicx}
\usepackage{dcolumn}
\usepackage{bm}
\usepackage{hyperref}
\usepackage{xcolor}
\usepackage{acronym}

\begin{document}

\preprint{APS/123-QED}

\title{All-sky search for short gravitational-wave bursts in the first part of the fourth LIGO-Virgo-KAGRA observing run}

\author{A.~G.~Abac\,\orcidlink{0000-0003-4786-2698}}
\affiliation{Max Planck Institute for Gravitational Physics (Albert Einstein Institute), D-14476 Potsdam, Germany}
\author{I.~Abouelfettouh}
\affiliation{LIGO Hanford Observatory, Richland, WA 99352, USA}
\author{F.~Acernese}
\affiliation{Dipartimento di Farmacia, Universit\`a di Salerno, I-84084 Fisciano, Salerno, Italy}
\affiliation{INFN, Sezione di Napoli, I-80126 Napoli, Italy}
\author{K.~Ackley\,\orcidlink{0000-0002-8648-0767}}
\affiliation{University of Warwick, Coventry CV4 7AL, United Kingdom}
\author{C.~Adamcewicz\,\orcidlink{0000-0001-5525-6255}}
\affiliation{OzGrav, School of Physics \& Astronomy, Monash University, Clayton 3800, Victoria, Australia}
\author{S.~Adhicary\,\orcidlink{0009-0004-2101-5428}}
\affiliation{The Pennsylvania State University, University Park, PA 16802, USA}
\author{D.~Adhikari}
\affiliation{Max Planck Institute for Gravitational Physics (Albert Einstein Institute), D-30167 Hannover, Germany}
\affiliation{Leibniz Universit\"{a}t Hannover, D-30167 Hannover, Germany}
\author{N.~Adhikari\,\orcidlink{0000-0002-4559-8427}}
\affiliation{University of Wisconsin-Milwaukee, Milwaukee, WI 53201, USA}
\author{R.~X.~Adhikari\,\orcidlink{0000-0002-5731-5076}}
\affiliation{LIGO Laboratory, California Institute of Technology, Pasadena, CA 91125, USA}
\author{V.~K.~Adkins}
\affiliation{Louisiana State University, Baton Rouge, LA 70803, USA}
% ...

\collaboration{The LIGO Scientific Collaboration, the Virgo Collaboration, and the KAGRA Collaboration}

\date{\today}

\begin{abstract}
We present a search for short-duration gravitational-wave transients
in data from the first eight months of Advanced LIGO-Virgo-KAGRA's
fourth observing run, denoted O4a.
We use four analyses which are sensitive to a wide range of potential signals
lasting up to a few seconds in the $16$--$4096$ Hz band.
Excluding binary black hole merger candidates that were already
identified by low-latency analyses, we find no statistically
significant evidence for other gravitational-wave transients.
We measure the sensitivity of the search for representative signals,
including sine-Gaussians, Gaussian pulses, and white-noise bursts with
different frequencies and durations, adopting a false alarm rate of 1 per 100 years as detection threshold.  
Depending on signal type, we find improvements
over previous searches by factors of $2$ to $10$ in terms of
sensitivity to strain amplitude and of 90\% confidence upper limit on the rate density of sources.
We also evaluate a variety of core-collapse supernova models and find
that, for some models, the search could have detected gravitational
waves from stellar core-collapse throughout the Milky Way.
Finally, we consider neutron star \textit{f}-modes associated with pulsar
glitches and find that, assuming a source similar to the Vela Pulsar,
the search could have detected a gravitational-wave signal from a
glitch with fractional frequency change as small as $\sim$$2$ to $6 \times 10^{-5}$
depending on the neutron star mass.

\end{abstract}

\maketitle

\section{\label{sec:Introduction}Introduction}

The Advanced LIGO~\citep{AdvLIGO2014} and Advanced Virgo~\citep{VIRGO:2014yos} detectors have proven the ability of large laser interferometers to detect and characterize gravitational waves.
Using data from the first three observing runs of these detectors, the LIGO Scientific Collaboration, Virgo Collaboration and KAGRA Collaboration (collectively \textit{LVK}) identified 90 transient signals most likely to be gravitational-wave (GW) events~\citep{GWTC-1,GWTC-2p1,GWTC-3}.
Other researchers have reported several additional GW event candidates in the publicly available data~\citep{2019ApJ...872..195N,2020ApJ...891..123N,2021ApJ...922...76N,2023ApJ...946...59N,2020PhRvD.101h3030V,2021PhRvD.104f3030Z,2022PhRvD.106d3009O,PhysRevD.111.024049,2023arXiv231206631W}.
All of the GW signals detected in those runs have matched the distinctive signature expected from compact binary coalescence (CBC) events.
A CBC arises from a co-orbiting pair of compact astrophysical objects,
specifically black holes or neutron stars,
that spiral inward and finally merge due to emission of gravitational radiation as predicted by the general theory of relativity.

Over 200 additional CBC candidates have been reported from low-latency analyses during the fourth LVK observing run~\citep{GraceDB}.
Many more CBC events will be detected as the LIGO and Virgo detectors improve and as the GW detector network grows with the maturation of the KAGRA detector~\citep{KAGRA:2020tym} and construction of a LIGO observatory in India~\citep{LIGOIndia2011,LIGOIndiaScience} as well as next-generation facilities currently being designed~\citep{ETScience2023,CEHorizonStudy}.
In parallel, the LVK continues to search for other types of GW signals.
For example, transient GWs---also called \textit{GW bursts}---can arise from the strong dynamics of mass and energy in
core-collapse supernovae~\citep{Janka2012,Gossan2016,SNGW_review,PhysRevD.104.102002,MezzacappaZanolin2024},
magnetars~\citep{NazariRoshan2020,Contopoulos2023,Kouvaris2025},
\textit{starquakes} of neutron stars~\citep{Watts2007,Giliberti2022},
nonlinear memory~\citep{Ebersold2020zah},
hyperbolic encounters between compact objects~\citep{Bini2023gaj}, and
cosmic strings~\citep{2000PhRvL..85.3761D,2001PhRvD..64f4008D,2005PhRvD..71f3510D}.
The time-frequency structure, the strength and the rate of occurrence of GW transient events depend on the source's internal microphysics,  formation channels and population models, which are not fully understood.
It is also possible that standard searches for CBC events, which use matched filtering with modeled waveforms as templates, may miss some binary mergers with very high masses, eccentric orbits and/or extreme spins, which are difficult to detect efficiently with template banks currently in use.
Therefore it is essential to search as broadly as possible for GW signals from potential sources, known or unknown.

Search algorithms which do not rely on having modeled waveforms have been developed and refined to meet this need.
Different methods are employed to search for generic short-duration GW burst signals (up to a few seconds long) and longer-duration transient signals over the full frequency range in which the detectors are highly sensitive.
Generic searches for short-duration~\citep{O3-allsky-short} and long-duration~\citep{O3-allsky-long} GW bursts in LIGO-Virgo data from the third observing run (O3) uncovered no significant candidates, other than already-known CBC events, despite having good sensitivity for a wide range of other potential signals.
These searches are complemented by dedicated analyses for GW signals from intermediate mass black hole (IMBH) mergers \cite{O3_imbh_2022} and compact binary systems with significantly eccentric orbits \cite{O3_eccentric_2024}, which obtain better sensitivity for those signal classes by making morphological assumptions specific to each.

In this paper we report on a search for short-duration GW bursts in data from the LIGO Hanford and LIGO Livingston Observatories (LHO and LLO, respectively) taken during the first eight months of the fourth observing run (O4), a period denoted O4a.
(A companion paper \cite{O4a-allsky-long} reports on a search for longer-duration transient GW signals.)
We give more details about the detectors, the O4a run period and data quality in section II.
In section III we describe the three search \textit{pipelines} used in this paper.
All three pipelines are based on the Coherent WaveBurst algorithm~\citep{klimenko2016method,drago2021coherent} but use different configurations, time-frequency transforms, and candidate event classification algorithms to obtain a broad coverage of the signal parameter space.

As we report in section IV,
this search finds many of the likely CBC signals that have already been reported publicly from low-latency analysis (and which will be verified and discussed in later LVK papers) but our goal here is to reveal any short-duration transient GW signal that is \emph{not} a CBC event.
We find no statistically significant candidate when the likely CBC signals are excluded.
In section V we evaluate and compare the sensitivities of the analyses using simulations which sample a wide range of potential signals in the detectors' sensitive frequency range.
The O4a run was only about half as long in duration as O3,
but due to lower noise in the LIGO detectors and some analysis improvements,
this early look at the O4 run covers a larger astrophysical volume-time exposure than O3 by a factor of 2 to $\sim$10, depending on the signal morphology.

In section VI we evaluate the sensitivity of this search to modeled signals from core-collapse supernovae and sudden excitations of neutron stars, providing astrophysical interpretation of our null search results.
This \textit{all-sky search} does not rely on observing a supernova light curve or neutron star pulse timing glitch, so it complements multi-messenger searches that are triggered by electromagnetic or neutrino observations, e.g.\ \cite{LVK-SN2023ixf}.
We end with some remarks about the ongoing O4 run and prospects for future detections.

\section{\label{sec:Run}Detectors and Data}

	\subsection{\label{sec:Detectors}Detectors Operating During O4a}

	  	\begin{figure*}
			\begin{center}
			\includegraphics[width=2.0\columnwidth]{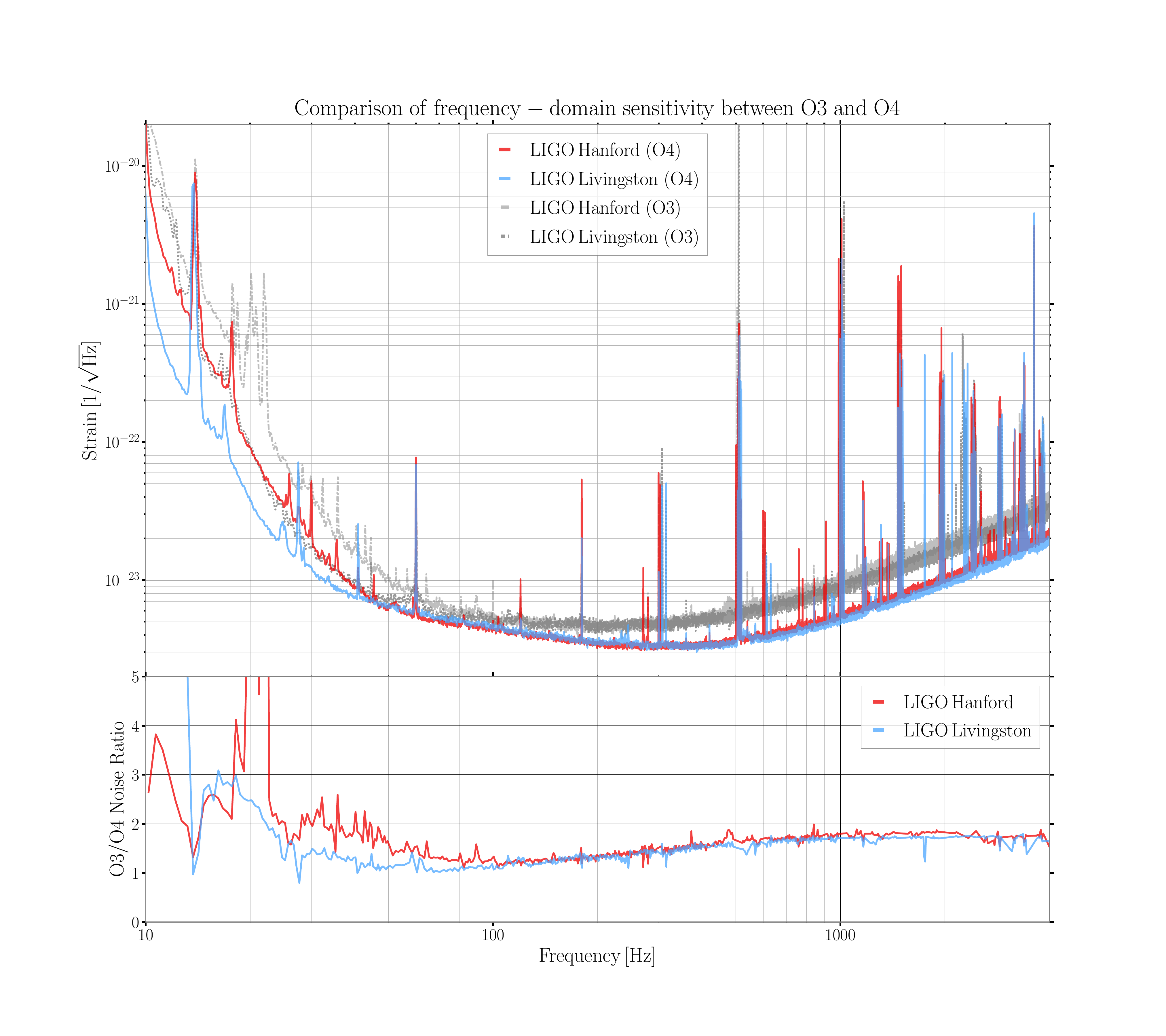}
                        \end{center}  
			\caption{A comparison of GW detector noise between observatories and observing runs. Top: amplitude spectral densities (ASDs) of the GW strain noise for the LHO (red) and LLO (blue) detectors in the O4 run compared to the O3 run (gray). Virgo's and KAGRA's strain sensitivities are not plotted. Bottom: the ratio between each LIGO detector's O3 and O4 noise ASD as a function of frequency. A larger value indicates a greater reduction in noise between O3 and O4. This ratio comparison is omitted near power line harmonics and near the thermally-driven ``violin'' resonance modes of optics suspensions above 250 Hz.}
			\label{fig:ligo_asds}
		\end{figure*}

The LIGO detectors underwent significant upgrades and commissioning between the conclusion of the O3 observing run on March 27, 2020 and the beginning of the O4 run on May 24, 2023~\cite{commissioning,detchar}.
Those efforts reduced the detector noise at almost all frequencies, as shown in Figure~\ref{fig:ligo_asds}.
The improvement at low frequencies is due to realizing frequency-dependent squeezing~\cite{sqzpaper} and reducing sources of environmental and technical noise, while the improvement at high frequencies is mainly due to increasing the laser power input into the interferometers to 60--75 W from 37 W in O3~\cite{commissioning}.
The reduced detector noise enables more sensitive searches for GW signals, and more stringent limits on source populations and GW emission mechanisms from searches that find no significant candidates.

In addition to the LIGO detectors, the KAGRA GW observatory participated in O4a from the beginning until June 20, 2023, at which point it was taken offline to conduct commissioning activities. KAGRA data are not used in this work due to its much lower sensitivity. The Virgo observatory was being upgraded during all of O4a. Virgo resumed joint observation with the LIGO detectors in April 2024 for the continuation of the O4 run.

	\subsection{\label{sec:O4aData}O4a Data}

O4a lasted from May 24, 2023 15:00 UTC to January 16, 2024 16:00 UTC. At least one LIGO facility was observing for 83.1\% of this period.
To carry out the search for GW signals, the four analyses described in section~\ref{sec:Analyses} are applied to coincident data
of high quality (see section~\ref{sec:DQ}) from both LIGO detectors.
Additionally, the analyses require a minimum of 200 seconds of contiguous observing-mode data and discard the first and last 4 or 10 seconds from each period of observing time in each detector. These restrictions reduce the dataset to 126.4 days of dual-coincident, high quality observing-mode data from the LIGO detectors in O4a, representing 53.3\% of the total duration of the O4a run.

The LIGO detectors employ photon calibrators to determine the observed GW strain based on fiducial displacements of the end test masses due to radiation pressure from an auxiliary laser system~\cite{2020CQGra..37v5008S,2021arXiv210700129S}. The frequency-dependent calibration uncertainty varies during the run and is assessed hour by hour.
At the beginning of O4a the amplitude uncertainty is typically $\sim$5\% or less for LHO and $\sim$3\% or less for LLO for frequencies below 1 kHz, rising to $\sim$10\% at 4 kHz. For data collected after June 13, 2023 the LLO calibration uncertainty improves, becoming typically $\sim$2\% at all frequencies up to 4 kHz. After July 27, 2023 the LHO calibration uncertainty improves, becoming typically $\sim$2\% up to 1 kHz, rising to $\sim$5\% at 4 kHz.

\subsection{\label{sec:DQ}Data Quality}

	\subsubsection{Data Quality Vetoes}

Detector data quality varies during observing-mode time and occasionally is poor; these periods can sometimes be identified and vetoed. 
In O4a, two categories of data quality (DQ) vetoes are applied~\cite{detchar}: category 1 vetoes flag time intervals with such severe issues that data taken during these times are not used by search pipelines, while category 2 vetoes represent times when the data quality is poor but certain time or frequency bands may still contain usable data.

In O4a, modeled searches for gravitational waves from compact binary coalescences exclude only category 1 vetoed times, while the analyses described in this work exclude the times of both category 1 and 2 vetoes, because the lack of a distinct morphology (like the ``chirp'' of a CBC) for unmodeled bursts complicates the prospect of distinguishing signals from noise.

Category 1 vetoes in O4a include periods when the observing mode state at an observatory was incorrectly activated (i.e., the detectors were not truly ready to observe) and periods of elevated motion in the detector suspensions that significantly degraded the calibration of LLO. Examples of category 2 vetoes used in this work include periods when large noise transients in the GW data were witnessed by voltage monitors in the electrostatic drive system used to control the position of the test masses; these transients are temporally correlated with instrumental artifacts (\textit{glitches}) in the GW data. Additional criteria for vetoing O4a data may be found in~\cite{detchar}. The total amount of observing-mode time vetoed in O4a was 1.66 days at LHO and 3.36 days at LLO.
	\subsubsection{\label{sec:OtherNoise}Other Transient Noise}

Despite the efforts to isolate the LIGO detectors from noise, they are still subject to occasional noise transients of both known and unknown origin. The most deleterious class of noise transients for this work are \emph{blip glitches}~\cite{blips}. Like many of the GW waveforms studied herein, blip glitches are short-duration, large-bandwidth transients. Their cause is uncertain. They are present in both LHO and LLO data, but at uncorrelated times. Most are not witnessed by auxiliary monitoring channels and therefore cannot be vetoed. According to Gravity Spy~\cite{gspyupdate}, blip glitches occurred at an average rate of 1.24 per hour at LHO and 0.87 per hour at LLO. This is a decrease from the O3 rate of 1.68 and 3.47 blip glitches per hour, respectively. Investigations to determine the cause of blip glitches have included searching for temporal correlations with vibrational transients along the LIGO beamtubes, temporal correlations with cosmic ray strikes, and correlations with environmental data monitors at each site. These investigations have so far not identified the cause of blip glitches in the LIGO detectors, although a small fraction of these glitches are contemporaneous with computer timing errors~\cite{o2o3detchar,detchar,blips}.

Especially at low frequencies, LIGO detector sensitivity can be degraded by stray light beams scattering off vibrating surfaces and interfering with the main beam, causing \emph{scattering glitches}~\cite{o3scatter}. Prior to O4a, a campaign was conducted to identify and mitigate several sources of scattered light noise~\cite{detchar}. While a direct comparison between the O3 and O4a searches is difficult to make due to differences in search pipeline configurations, the incidence of large signal outliers below 128~Hz was reduced in O4a compared to O3. This is believed to be a result of improved isolation from scattered light noise.

\section{\label{sec:Analyses}Search Algorithms, Pipelines, and Analyses}

\acrodef{2G}[2G+XGB]{\emph{}}
\acrodef{XP}[XP+XGB]{\emph{}}
\acrodef{GMM}[2G+GMM]{\emph{}}

\acrodef{cwb2G}[cWB-2G]{\textit{second-generation}}
\acrodef{cwbXP}[cWB-XP]{\textit{cross-power}}

The search algorithms used in this paper are based on the open-source software Coherent WaveBurst (cWB) \cite{klimenko2016method,drago2021coherent,cwb},
which is designed to operate without a specific signal model.
cWB begins by computing multi-resolution time-frequency (TF) representations of the strain data from the GW detector network \cite{necula2012transient, klimenko2022wavescan}. 
The multi-resolution approach is essential to cover the wide parameter space of potential signals with durations ranging from milliseconds to a few seconds in a frequency range from tens of Hz up to a few kHz.
cWB selects the most energetic TF regions, which are referred to as \textit{triggers} if they pass internal pipeline thresholds, and evaluates how potential GW signals would appear in the network of detectors.
For each trigger, a likelihood ratio statistic is maximized over all sky directions \cite{klimenko2005constraint} enabling cWB to reconstruct the source sky localization, the signal waveforms, and morphological properties of the signals such as the duration, the central frequency, etc. 
The ranking statistic $\eta$ is essentially the signal-to-noise ratio (SNR) of the trigger, estimated from the coherent response of the network, with a penalty factor related to the incoherent energy residuals in the detectors. The trigger significance is measured in terms of inverse false alarm rate (IFAR), computed by comparing the ranking statistic of each trigger with a \textit{background} distribution. The latter is built by repeating the cWB algorithm on time-shifted data, shifting one detector with respect to the other detector by an amount that breaks any coherence from astrophysical signals. Previous versions of cWB were used to search and reconstruct GW transients in past LVK observing runs  (O1 \cite{allskyO1, allskyO1long}, O2 \cite{allskyO2, allskyO2long}, and O3 \cite{O3-allsky-short, O3-allsky-long}).\par 
To maximize the discovery potential of our search, we employ two updated versions of cWB to generate triggers, 
with shorthand names derived from \ac{cwb2G} and \ac{cwbXP} development paths, presented in Section\ \ref{sec:processing}. 
After the generation of the triggers, two different machine-learning procedures---a decision-tree algorithm called XGBoost and a Gaussian mixture model (GMM)---classify the triggers and enhance the separation between possible astrophysical transients and detector noises, as described in Section\ \ref{sec:post}.\par
\begin{figure}
  \begin{center}
  \includegraphics[width=.5\textwidth]{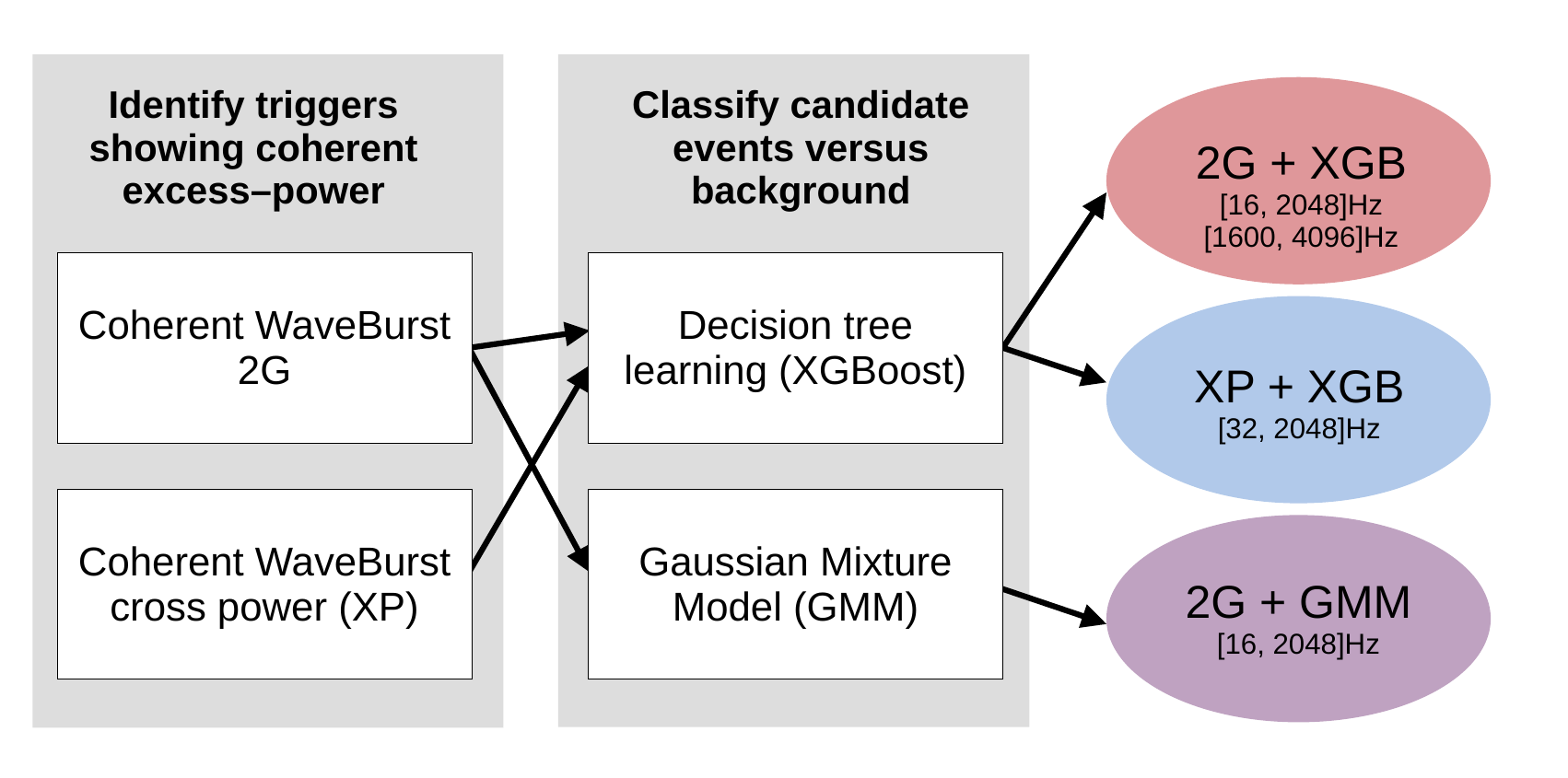}
  \end{center}
	\caption{The GW burst search reported in this paper uses two algorithms to identify triggers, i.e.\ coherent excess of power, described in Section\ \ref{sec:processing}, and two machine-learning classification algorithms to suppress background, described in Section\ \ref{sec:post}. These algorithm components are combined to construct three \textit{pipelines}: \ac{2G}, \ac{XP}, and \ac{GMM}. The \ac{2G} pipeline is used in two \textit{analyses}, one for low-frequency signals and the other for high-frequency signals, while each of the other pipelines is used in a single analysis covering a similar low-frequency range.
The complementary aspects of these four analyses are discussed in the text.
  }
        \label{fig:flowchart}
\end{figure}

The total number of search pipelines applied to the O4a dataset is three, illustrated in Figure \ref{fig:flowchart}: two use the triggers generated by the \ac{cwb2G} algorithm and apply XGBoost\ac{2G} and GMM\ac{GMM} classification algorithms, respectively, while the third generates triggers with the \ac{cwbXP} algorithm and then applies XGBoost\ac{XP}. 
All three pipelines operate in the frequency band where the detectors are most sensitive, up to 2048 Hz, aiming to maximize their combined detection efficiency in this parameter space. Although the pipelines perform similarly on average, as shown in Section \ref{sec:Sensitivity}, their detection efficiency varies across different waveform morphologies, signal amplitudes, and polarizations. They also respond differently to non-stationary noise. Our simulation studies reveal that each pipeline detects some signals that the others miss.
Additionally, the \ac{2G} pipeline performs a dedicated analysis for high-frequency signals, so a total of four \emph{analyses} make up the search reported in this paper. \par
The \ac{2G} and \ac{XP} pipelines have also been used to search for generic GW transients in low latency while the GW detectors were collecting data. The main difference between the configurations used for low-latency analyses and the offline analyses presented here is in the training of the machine-learning algorithms used to classify candidate events versus transient noise (Section \ref{sec:xgb}).

\subsection{Identify triggers showing coherent excess--power}
\label{sec:processing}
The Coherent WaveBurst algorithm has been successfully employed in previous GW burst searches, e.g.\ \cite{O3-allsky-short}. In the following we provide details about the two upgraded versions of cWB employed to generate triggers from the GW strain data.
\subsubsection{\label{sec:Pipe1}Coherent WaveBurst 2G}
The \ac{cwb2G} algorithm computes the TF representation with the Wilson-Daubechies-Meyer wavelet transform \cite{necula2012transient}, using multiple TF resolutions to adapt to different signal morphologies. The \ac{cwb2G} algorithm improves on the selection of TF pixels that form trigger candidates to improve the sensitivity to low SNR events. After selection, TF pixels are aggregated if they are separated by time and frequency intervals of less than 0.2 s and 640 Hz; this improves the collection of energy for transient events with complex time-frequency structure that appears as separate clusters of TF pixels.
\ac{cwb2G} is used in two configurations with different sets of TF resolutions to perform low-frequency (16--2028 Hz) and high-frequency (512--4096) analyses. High-frequency analysis triggers with central frequency below 1600 Hz are removed, leaving a 448 Hz overlap with the low-frequency analysis to ensure good coverage of wide-band GW events at intermediate frequencies. A trials factor of 2 is applied to the IFAR of triggers in this overlap region, accounting for the fact that they could be found by both low- and high-frequency analyses.

\subsubsection{\label{sec:Pipe2}Coherent WaveBurst Cross Power}
The \ac{cwbXP} algorithm implements two major changes with respect to the cWB version used in the O3 search \cite{O3-allsky-short}. First, the Wilson-Daubechies-Meyer wavelet transform is replaced with the multi-resolution \emph{WaveScan} transform based on Gabor wavelets \cite{klimenko2022wavescan} in order to reduce temporal and spectral leakage in the TF representation.
For O4a, \ac{cwbXP} analyzes the frequency range between 32 Hz and 2048~Hz.
After selection, the initial TF clusters are aggregated if they are separated by time and frequency intervals of less than $0.23~\mathrm{s}$ and $64~\mathrm{Hz}$, respectively.
A coherent SNR$>7$ threshold is applied to aggregated clusters, resulting in a trigger rate of $O(0.01\,\mathrm{Hz})$.
Second, in addition to the excess-power statistic used previously in cWB, \ac{cwbXP} applies a cross-power statistic \cite{klimenko2022wavescan} to identify transient events and suppress the noise transient rate. The cross-power amplitude 
of each TF data sample (or \emph{WaveScan} pixel) is maximized over all possible time-of-flight delays of a GW signal in the detector network. Both statistics follow a predictable half-normal distribution with unit variance expected for quasi-stationary detector noise and are described in Refs.\ \cite{klimenko2022wavescan, PhysRevD.111.023054}.

\subsection{Classify candidate events versus background}
\label{sec:post}
The output rate of both \ac{cwb2G} and \ac{cwbXP} algorithms is dominated by triggers originating from non-stationary detector noise. 
To increase the confidence of detections, we have replaced the manual tuning of selection thresholds, used through O3~\cite{O3-allsky-short}, with procedures exploiting two machine-learning algorithms.
\subsubsection{Decision tree learning with XGBoost}
\label{sec:xgb}
XGBoost is a supervised decision tree learning algorithm \cite{chen2016xgboost} employed to classify \ac{cwb2G} and \ac{cwbXP} triggers between a noise class and a signal class. XGBoost receives as input tabular values from the two labelled classes and, during the training process, builds a multi-dimensional model which outputs a number ranging from 0, for triggers supposed to belong to the noise class, to 1 for the signal class.\par
The training dataset uses unmodified background triggers for the noise class, while the signal class is populated with generic white-noise burst (WNB) injections selected to cover the full parameter space of the search \cite{cwb_xgb_allsky}. The tabular values used to train the model are a subset of summary features (10 for \ac{2G} and 16 for \ac{XP}) that do not depend directly on the morphological characteristics of anticipated signals, in order to maintain the unmodeled nature of the search. The selected summary statistics describe the degree of correlation of the trigger in the detector network, the energy that is coherent and not-coherent in the network, the energy in each TF resolution, and attributes measuring the likeness to known single-cycle glitches. The detailed list of summary features used by \ac{2G} can be found in Ref.\ \cite{cwb_xgb_allsky}. \ac{XP} includes in the training also summary statistics quantifying the cross-power and the network alignment factor \cite{PhysRevD.111.023054}.
In the low-frequency region, the WNB injections used to train the XGBoost model have duration uniformly distributed in [0.001,0.5]~s, central frequency in [24,1696]~Hz, and bandwidth in [10,800]~Hz. In the high-frequency region (512--4096~Hz), the WNB injections used to train the XGBoost model have duration uniformly distributed in [0.0001,0.5]~s, central frequency in [1200,3400]~Hz, and bandwidth in [100,1600]~Hz.\par
The \ac{2G} and \ac{XP} low-latency analyses update their XGBoost model when the properties of the noise distribution change substantially. The WNB simulations and background triggers used to train the models are acquired continuously during the observing run.

For the offline search presented here,
\ac{cwb2G} triggers are processed by XGBoost models trained on data from limited periods of time. The full dataset is divided into data segments, referred to as chunks, of about 9 days each. 
For each chunk, an XGBoost model is trained on background and WNB injection triggers acquired on the two adjacent chunks (both sides, apart from end chunks) plus an additional 10\% of the same chunk's triggers. The remaining 90\% background triggers of the chunk are used to estimate the statistical significance of candidates. This training procedure adapts well to non-stationary properties of the data, and at the same time prevents model overtuning. Once trained, the XGBoost output $W_{\mathrm{XGB}}$ is stretched monotonically to enhance sensitivity for events with $W_{\mathrm{XGB}}$ close to unity and used as the ranking statistic for \ac{cwb2G} triggers.\par

In contrast, the classification stage for \ac{cwbXP} uses a single XGBoost model trained on data from the entire run along with the simulated WNBs. Approximately 20\% of the background triggers are used for training, while the remaining background triggers are used to estimate the statistical significance of candidates.
The ranking statistic is the one described in \cite{cwb_xgb_allsky, cwb_xgb_BBH}:
\begin{equation}\label{eq:XP}
\eta_\mathrm{r} = \eta_\mathrm{0}\cdot W'_{\mathrm{XGB}},
\end{equation}
where $W'_{\mathrm{XGB}}$ is computed from the output of the XGBoost model \cite{xgb_optimization}, including a monotonic stretch and a correction to mitigate the impact of loud blip-like glitches. It is used as a multiplicative factor to re-weight $\eta_\mathrm{0}$, the cWB detection statistic that is based on the coherent energy~\cite{xgb_optimization}.

\subsubsection{Gaussian mixture model}
An alternative supervised machine-learning approach, GMM, is applied to \ac{cwb2G} triggers in the \ac{GMM} pipeline. GMM learns to model multi-dimensional distributions as superpositions of Gaussians. The application of GMM to enhance GW burst searches is described in Refs.\ \cite{gayathri2020_gmmO1, lopez2022_gmmO2, PhysRevD.110.083032}. GMM constructs two distinct models, one for the noise and one for the signal, from a subset of 11 \ac{cwb2G} trigger summary statistics which undergo re-parameterization in order to have desired Gaussian behaviour. The ranking statistic $\eta$ 
is defined by a log-likelihood ratio $\eta = W_s - W_n$, where $W_s$ and $W_n$ are the maximum log-likelihood statistics derived from the signal and noise models, respectively. 
The training dataset includes background triggers and WNB injections as described above for the XGBoost algorithm. The \ac{GMM} pipeline covers the frequency range [16,2048]~Hz and analyzes data segments formed from two adjacent chunks at a time. For each segment, GMM models first use 5\% of the background triggers and 20\% of WNB simulations to select the optimal number of Gaussians required to efficiently model the data, then 25\% of the background triggers and 80\% of the WNB simulations to train the models. The remaining 70\% of the background triggers are used to evaluate the statistical significance of candidates.

\section{\label{sec:Results}Search Results}

In this section we report the results of the four analyses performed using the search pipelines presented above.

While these search methods are designed to explore a wide range of possible signals, the low-frequency analyses in particular are expected to detect CBC events since most of the signal power of a CBC is at frequencies below 1 kHz.
However, being agnostic on the signal morphology, GW burst searches have a higher false alarm rate than template-based methods so their detection capability for CBC signals is somewhat lower, or comparable in the case of IMBH binaries \cite{O3_imbh_2022}.

\begin{figure*}
\begin{minipage}[b]{0.5\linewidth}
	\centering
  \includegraphics[width=.92\textwidth]{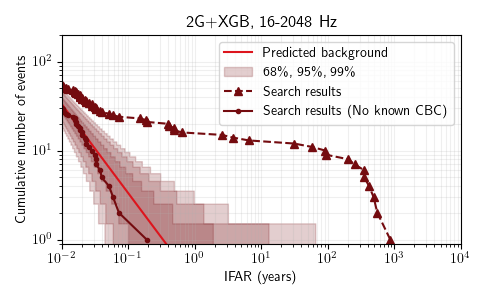}
\end{minipage}%
\begin{minipage}[b]{.5\linewidth}
  \centering
  \includegraphics[width=.92\textwidth]{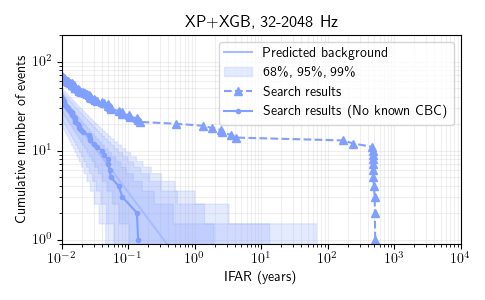}
\end{minipage}%

\begin{minipage}[b]{0.5\linewidth}
        \centering
  \includegraphics[width=.92\textwidth]{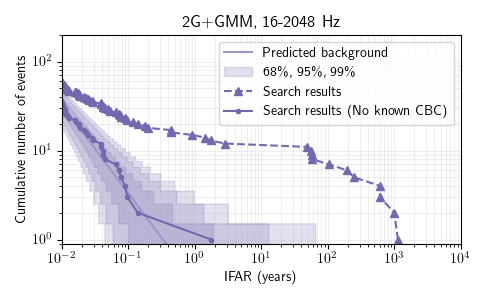}
\end{minipage}%
\begin{minipage}[b]{.5\linewidth}
  \centering
  \includegraphics[width=.92\textwidth]{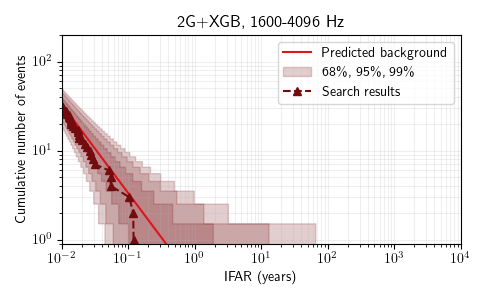}
\end{minipage}%
		\caption{Cumulative number of candidate events versus IFAR found in the low-frequency analyses by \ac{2G} (top left), \ac{XP}  (top right), \ac{GMM} (bottom left), and in the high-frequency analysis by \ac{2G} (bottom right).  Triangular points indicate the results for all data, while circular points show candidates remaining after CBC candidate signals found in low latency have been excised. The solid line shows the expected mean value of the background, while the shaded regions indicate 68\%, 95\%, and 99\% Poisson uncertainty regions.}
\label{fig:zerolag}
\end{figure*}

The results of the four analyses performed are shown in Figure \ref{fig:zerolag}.
Each panel of that figure shows the cumulative set of event candidates ordered by IFAR and compared to the straight-line expectation for false alarms.
The triangles connected with dashed lines show a substantial excess in each low-frequency analysis, representing detected GW candidate events. \ac{XP} accumulates about 500 years of background for each chunk, so there is a sharp decrease at IFAR around 500 years in the upper right panel in Fig.\ \ref{fig:zerolag}. In contrast, \ac{2G} and \ac{GMM} accumulate different amounts of background statistics for each chunk, so there is a less-sharp edge to the cumulative number of GW candidate events in the two left panels.\par
The circles connected with solid lines show the distribution after all CBC candidates, identified by low-latency analyses \cite{klimenko2016method,Tsukada:2023edh, Ewing:2023qqe,DalCanton:2020vpm,Chu:2020pjv, Allene2025} with an IFAR larger than one month and reported in public alerts (including S231123cg which was later confirmed as GW231123 \cite{GW231123preprint}), have been excised \footnote{The CBC candidates excised from this search will be published in the next update to the LVK Gravitational-Wave Transient Catalog (GWTC) and online in the Gravitational Wave Open Science Center, \url{http://www.gwosc.org}.}.
No significant individual event remains after that excision, nor any significant cumulative excess in the IFAR distributions. In the high-frequency analysis, where we do not expect to detect CBC signals, the search results are consistent with the expected distribution of false alarms.\par 

The most significant candidate event from the low-frequency analyses is found on 2023-08-30 at 05:07:03 UTC by the \ac{GMM} pipeline with IFAR of 1.75 years.
The same trigger is identified by \ac{2G} with IFAR of
0.059 year, while it is not identified by \ac{XP}. This candidate has an SNR of 19 concentrated at about 930~Hz. Using the method described in \cite{Helmling-Cornell:2023wqe}, we checked for the presence of linear coupling between the network of environmental monitoring sensors \cite{nguyen2021environmental} and the GW strain data at each LIGO observatory. No vibrational or magnetic transients witnessed by these sensors contaminated the GW data at either site around the time of this candidate event.
Although we cannot reject this candidate on the basis of data quality, its statistical significance is not high enough to be claimed as a GW event.\par
	The second most significant low-frequency event candidate has a low significance (IFAR $\le 0.2$ year in all analyses), so it was not investigated further. \par
	The most significant candidate of the \ac{2G} high-frequency analysis occurs on 2023-12-23 at 18:25:36 UTC with IFAR of 0.12 year. This candidate has SNR 14.7, central frequency of 3146~Hz and duration of 0.023~s. We did not investigate this candidate further due to its low significance.\par

To assess the complementarity of the different algorithms, we consider candidate events found by each low-frequency analysis with IFAR $>1$ day. As before, we exclude known CBC candidates, so that the remaining low-significance candidates are most likely due to detector noise fluctuations. 
There are 316 such candidates, of which 106 are found by \ac{2G}, 129 by \ac{XP}, and 137 by \ac{GMM}. Comparing their times, we find that only 10 of these candidates are found by all three pipelines, with an additional 4 found by both \ac{2G} and \ac{XP}, 5 by \ac{XP} and \ac{GMM}, and 27 by \ac{2G} and \ac{GMM}. The fact that the majority of the low-significance candidates are found by only one pipeline means that the three pipelines are fairly independent in their response to detector noise fluctuations.

\section{\label{sec:Sensitivity}Search Sensitivity and Rate Limits}

As in previous iterations of this search, we assess the sensitivity of our analyses to a variety of ad-hoc signals that cover broad ranges of frequency, duration, and bandwidth. Simulated waveforms are generated with chosen sky directions and injected into data from the network of GW detectors with the proper polarization and time delays to mimic a true GW signal.  The amplitude distribution is chosen such that for each waveform and each analysis the detection efficiency goes from around 0 to 100\%. The distribution in sky direction is uniform in solid angle over the whole sky. The waveform families used are Gaussian pulses (GA) with varying durations parametrized by the Gaussian standard deviation of time $\tau_{GA}$, sine-Gaussians (SG) with varying central frequency $f_0$ and quality factor $Q$ proportional to the signal duration, and band-limited white-noise bursts parameterized by a central frequency $f_0$, bandwidth $\Delta f$, and duration $\tau_\mathrm{WNB}$. Further description of these waveforms can be found in \cite{S6_Burst}. The amplitude is characterized by a quantity called the root-sum-squared strain amplitude, $h_\mathrm{rss} = \sqrt{\int_{-\infty}^{\infty} \left(h_{+}^{2} + h_{\times}^2\right) dt }$, where $h_+$ and $h_{\times}$ are GW polarizations in TT-gauge. Analysis is done using data plus simulated signals of each waveform type with nine different amplitudes, generated in terms of $h_\mathrm{rss}$ following the rule $h_\mathrm{rss} = (\sqrt{3})^{N} \times 5 \times 10^{-23} \,\mathrm{Hz}^{-\frac{1}{2}}$ where $N$ is an integer ranging from 0 to 8.

The polarization distribution is waveform dependent.
For the GA waveforms it is fixed to linear polarization mimicking linear motion emission.
For SG waveforms, we sample different elliptical polarizations determined by an inclination angle assuming a rotating source, which leads to polarization modes 
\begin{equation}
\begin{aligned}
h_{+}(t) &= \frac{1}{2}(1 + \cos^2 \iota) A_{+}(t) ,\\
h_{\times}(t) &= \cos(\iota) A_{\times}(t)
\end{aligned}
\end{equation}
where $\iota$ is the inclination angle (effectively the angle between the rotation axis and the line of sight) and $A_{+,\times}(t)$ is the time-dependent amplitude given by the SG (and analogous cosine-Gaussian) waveform.  The polarization distribution for SG follows a uniform distribution of $\cos(\iota)$. For WNB signals, the polarization distribution is kept random, representing isotropic emission.

We use a characteristic amplitude, $h_\mathrm{rss}^{50\%}$, as a measure of detection sensitivity.  $h_\mathrm{rss}^{50\%}$ is the injected $h_\mathrm{rss}$ for which the signal recovery efficiency reaches 50\%, which we determine by fitting the efficiency tabulated at the discrete values of $h_\mathrm{rss}$ used in the simulation and interpolating from the fit.  We report the results for all the waveforms at IFAR $\geq 100$ years in Table \ref{Tab.O3hrss50}. This IFAR detection threshold corresponds to p-value of $3 \times 10^{-3}$ which is equivalent to a $\sim$3 sigma detection in the O4a dataset.
Compared to the corresponding O3 results, we see improvements for all the morphologies. We note that the calibration uncertainties (Sec.\ \ref{sec:O4aData}) are not taken into account here. In response to the better low-frequency sensitivity of the detectors during O4 we have introduced lower-frequency SG injections at 36 and 48 Hz.
The largest improvement in $h_\mathrm{rss}^{50\%}$ is for the GA waveforms, for which the O4a sensitivity is at least 6 times better than O3. This is attributed to both improvements in the search algorithms and better techniques to isolate the very short-duration and loud glitches, which anyway occur with a lower rate in O4a as compared to O3, as noted in Sec.\ \ref{sec:OtherNoise}. For signals below 2 kHz the results are similar for all three low-frequency analyses, meaning that although the pipelines respond fairly differently to transient noise outliers (section \ref{sec:Results}), the overall sensitivity of these pipelines to a broad variety of waveforms is similar.
The \ac{GMM} pipeline performs best for GA waveforms and $h_\mathrm{rss}^{50\%}$ is typically within 15\% to 50\% of the other two pipelines for SG and WNB waveforms.
Beyond the fact that $h_\mathrm{rss}^{50\%}$ values are similar, each pipeline recovers some number of injected signals that are not identified by the others.

We can define an astrophysically meaningful quantity to interpret $h_\mathrm{rss}^{50\%}$ in terms of the energy emitted by a hypothetical source. Consider a standard candle source at fixed distance of $r_0$ radiating GWs in a narrow frequency range around a central frequency $f_0$.  The energy radiated by a rotating system like a circular binary (modeled here as an elliptically polarized SG) will be \cite{sutton}
\begin{equation} \label{evf}
    E_{\mathrm{GW}}^{\mathrm{rot}} = \frac{2}{5}\frac{\pi^2 c^3}{G} r_0^2 f_0^2 h_{\mathrm{rss}}^2 \,
\end{equation}
where $G$ is the gravitational constant and $c$ is the speed of light. The narrow-band approximation causes very low biases for our SG waveforms ($\leq 3\%$). We use each $h_\mathrm{rss}^{50\%}$ in the above equation to obtain this energy as a function of central frequency for $r_0 = 10\,\mathrm{kpc}$ as shown in Table \ref{Table:Evf}.  As compared to the previous observing run we improve by a factor of 2.5 to 6 across the different frequencies. These improvements result from both the better sensitivity of the detectors and improvements in the search methods.

We can also compute a limit on the rate per unit volume for a hypothetical population of sources producing these waveforms.
The generic source is assumed to emit $1\,M_{\odot}c^2$ of GW energy where $M_{\odot}$ is the mass of the Sun. Using equation \ref{evf} with $E_{\mathrm{GW}} = 1\,M_{\odot}c^2$, the relation between the average $h_\mathrm{rss}$ at the source at distance $r$ and on Earth is then $hr=h_0r_0$, where we have dropped the subscript $\mathrm{rss}$ for brevity. Assuming an intrinsic event rate of $R$ per unit volume per time, the expected number of detected events for an observing time $T$ is
\begin{equation}
\begin{aligned}
N_{\mathrm{det}} &= 4 \pi R T \int_0^{\infty} dr \, r^2 \, \epsilon(r) \\
&= 4 \pi R T (h_0 r_0)^3 \int_0^{\infty} dh \, h^{-4} \, \epsilon(h)
\end{aligned}
\end{equation}
where $\epsilon(h)$ is the efficiency as a function of amplitude.

\begin{figure}[t!]
  \begin{center}
  \includegraphics[width=1.0\columnwidth]{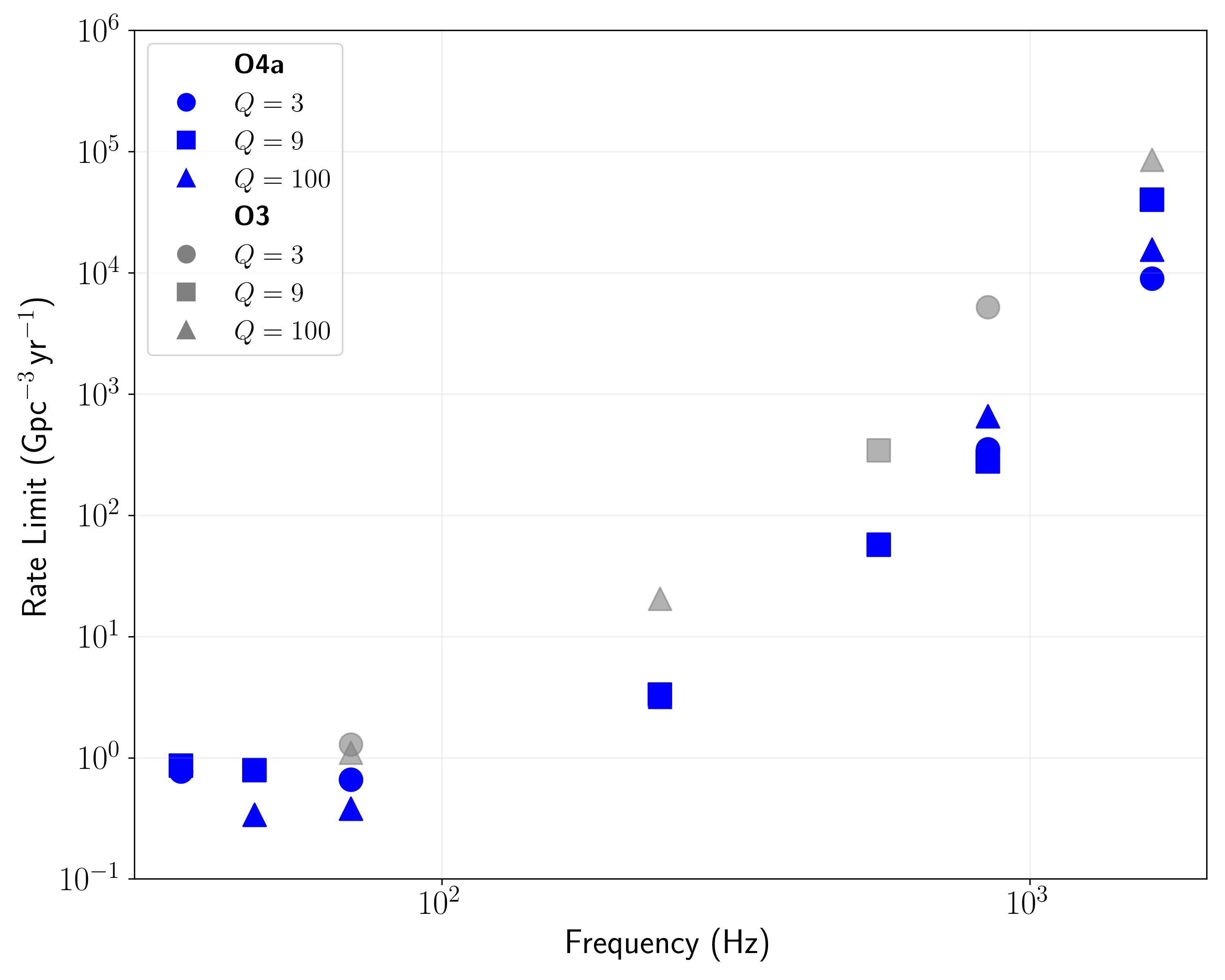}
  \end{center}
  \caption{Rate density upper limits (at 90\% confidence) in terms of Gpc$^{-3}$yr$^{-1}$ for elliptical sine-Gaussian injections, assuming standard candle sources emitting $1 M_\odot c^2$ of GW energy, for the best among the three low-frequency analyses in the case of O4a. Upper limits for waveforms with different $Q$ values are plotted versus frequency. Gray markers show the O3 results for the SG waveforms listed for O3 in table \ref{Tab.O3hrss50}.}
    \label{Fig:Rate-limit}
\end{figure}

In the absence of a detection, the rate limit at 90\% confidence is
\begin{equation}
    \mathcal{R}_{90\%} = \frac{2.3}{4 \pi T (h_0 r_0)^3 \int_0^{\infty} dh \, h^{-4} \, \epsilon(h)} \,.
\end{equation}
We use the SG waveforms given in Table \ref{Table:Evf} for these computations and fix the threshold at IFAR $\geq 100$ years. Results from the best of the three low-frequency analyses, on a waveform-by-waveform basis, are shown in Figure \ref{Fig:Rate-limit}. While the observing time of O4a was approximately half that of O3, the O4a analysis results in lower rate limits for elliptically polarized SG waveforms across all frequencies by a factor of 2 to $\sim$10. For instance, for the SG with $f_0 = 70$ Hz and $Q=100$, the O3 rate limit was 1.08 $\text{Gpc}^{-3} \text{yr}^{-1}$ while with O4a we obtain 0.43 $\text{Gpc}^{-3} \text{yr}^{-1}$.

\begin{table*}[!htb]
\begin{center}
\scriptsize
\begin{tabular}{ c c c c c }
\hline
\hline
   & \multicolumn{4}{c }{ 50\% efficiency $h_\text{rss} (\times 10^{-22}$ $\text{Hz}^{-1/2}$)} \\
\hline
\textbf{Morphology} & \textbf{\ac{2G}}  & \textbf{\ac{GMM}} & \textbf{\ac{XP}} & \textbf{O3} \\
\hline
\textbf{Gaussian pulses (linear)}\\
$\tau_{\mathrm{GA}} = 0.1$ ms & 1.9 & 1.6 & 1.6 & 12.6 \\
$\tau_{\mathrm{GA}} = 2.5$ ms & 1.7 & 1.9 & 2.0 & 16.7 \\
\textbf{Sine-Gaussian wavelets (elliptical)}\\
$f_0 = 36$ Hz, $Q = 3$ & 3.6 & 5.7 & 3.7 & - \\
$f_0 = 36$ Hz, $Q = 9$ & 4.6 & 4.5 & 3.9 & - \\
$f_0 = 48$ Hz, $Q = 9$ & 2.2 & 3.1 & 2.1 & - \\
$f_0 = 48$ Hz, $Q = 100$ & 3.4 & 3.6 & 4.3 & - \\
$f_0 = 70$ Hz, $Q = 3$ & 1.4 & 1.9 & 1.5 & (2.9) \\
$f_0 = 70$ Hz, $Q = 100$ & 1.8 & 2.2 & 1.7 & (2.7) \\ 
$f_0 = 235$ Hz, $Q = 9$ & 0.6 & 0.8 & 0.6 & - \\
$f_0 = 235$ Hz, $Q = 100$ & 0.9 & 1.1 & 0.9 & (2.0) \\
$f_0 = 554$ Hz, $Q = 9$ & 0.9 & 1.3 & 0.9 & (2.0) \\
$f_0 = 849$ Hz, $Q = 3$ & 1.4 & 1.8 & 1.4 & (3.6) \\
$f_0 = 849$ Hz, $Q = 9$ & 1.2 & 1.6 & 1.1 & - \\ 
$f_0 = 849$ Hz, $Q = 100$ & 1.1 & 1.3 & 1.1 & - \\ 
$f_0 = 1304$ Hz, $Q = 9$& 1.7 & 2.3 & 1.7 & (3.6) \\
$f_0 = 1615$ Hz, $Q = 3$ & 2.9 & 3.1 & 2.9 & - \\
$f_0 = 1615$ Hz, $Q = 9$ & 2.1 & 2.9 & 2.2 & - \\
$f_0 = 1615$ Hz, $Q = 100$ & 1.8 & 2.0 & 1.8 & (4.3) \\
$f_0 = 2000$ Hz, $Q = 3$ & 2.7 & - & - & (6.0) \\
$f_0 = 2477$ Hz, $Q = 9$ & 3.0 & - & - & (7.2) \\
$f_0 = 3067$ Hz, $Q = 3$ & 4.3 & - & - & (9.8) \\
\textbf{White-noise bursts}\\
$f_0 = 100$ Hz, $\Delta f = 100$ Hz, $\tau_{\mathrm{WNB}} = 0.1$ s & 0.6 & 0.8 & 0.6 & 1.0 \\
$f_0 = 250$ Hz, $\Delta f = 100$ Hz, $\tau_{\mathrm{WNB}} = 0.1$ s & 0.6 & 0.7 & 0.6 & 1.0 \\
$f_0 = 750$ Hz, $\Delta f = 100$ Hz, $\tau_{\mathrm{WNB}} = 0.1$ s & 0.8 & 0.9 & 0.8 & 1.5 \\
\hline
\hline
\end{tabular}
\end{center}
\caption{The $h_\text{rss}$ values (in units of $10^{-22}$ $\text{Hz}^{-1/2}$) for which 50\% detection efficiency is achieved with an IFAR threshold of 100 years for each of the injected signal morphologies. 
The SG waveforms reported in this table have elliptical polarization.
``-'' denotes waveforms not evaluated for a given pipeline. The SG results with central frequency of 2000 Hz and higher are from the high-frequency analysis. We show the corresponding results for elliptically polarized SG waveforms for O3 in parentheses. Note that Table I in the O3 paper \citep{O3-allsky-short} reported values for circularly polarized SG.}
\label{Tab.O3hrss50}
\end{table*}

\begin{table*}[!htb]
\begin{center}
\scriptsize
\begin{tabular}{ c c c c c}
\hline
\hline
\textbf{Sine-Gaussian parameters} & \multicolumn{3}{c}{$E_{\mathrm{GW}}$ (in units of $10^{-10} M_\odot c^2$)} \\
\hline
	& \textbf{\ac{2G}} & \textbf{\ac{GMM}} & \textbf{\ac{XP}} & \textbf{O3} \\
\hline
$f_0 =  36$ Hz, $Q=  3$ &     1.4 &   3.6 &   1.5 &    -  \\
$f_0 =  36$ Hz, $Q=  9$ &     2.3 &   2.2 &   1.7 &    -  \\
$f_0 =  48$ Hz, $Q=  9$ &     0.9 &   1.9 &   0.9 &    -  \\
$f_0 =  48$ Hz, $Q=100$ &     2.3 &   2.5 &   3.6 &    -  \\
$f_0 =  70$ Hz, $Q=  3$ &     0.8 &   1.5 &   0.9 &   3.5 \\
$f_0 =  70$ Hz, $Q=100$ &     1.3 &   2.0 &   1.2 &   3.0 \\
$f_0 = 235$ Hz, $Q=  9$ &     1.7 &   3.0 &   1.7 &    -  \\
$f_0 = 235$ Hz, $Q=100$ &     3.8 &   5.7 &   3.8 &  18.8 \\
$f_0 = 554$ Hz, $Q=  9$ &      21 &    44 &    21 &   104 \\
$f_0 = 849$ Hz, $Q=  3$ &     120 &   200 &   120 &   790 \\
$f_0 = 849$ Hz, $Q=  9$ &      88 &   157 &    74 &    -  \\
$f_0 = 849$ Hz, $Q=100$ &      74 &   104 &    74 &    -  \\
$f_0 =1304$ Hz, $Q=  9$ &     420 &   760 &   420 &  1870 \\
$f_0 =1615$ Hz, $Q=  3$ &    1900 &  2100 &  1900 &    -  \\
$f_0 =1615$ Hz, $Q=  9$ &     980 &  1860 &  1070 &    -  \\
$f_0 =1615$ Hz, $Q=100$ &     720 &   890 &   720 &  4100 \\
$f_0 =2000$ Hz, $Q=  3$ &    2500 &    -  &    -  & 12200 \\
$f_0 =2477$ Hz, $Q=  9$ &    4700 &    -  &    -  & 27000 \\
$f_0 =3067$ Hz, $Q=  3$ &   15000 &    -  &    -  & 77000 \\ 
\hline
\hline
\end{tabular}
\end{center}
\caption{GW emitted energy $E_{\mathrm{GW}}$ in units of $10^{-10} M_\odot c^2$ that corresponds to 50\% detection efficiency at IFAR $\geq $ 100 years, assuming a source distance $r_0=10\,\mathrm{kpc}$.  We present the results for different SG elliptical waveforms with varying  central frequency  $f_0$ and quality factor $Q$ for each pipeline used in O4 and also for the waveforms published for O3.  ``-'' denotes waveforms not evaluated for a given pipeline or not for O3.}
\label{Table:Evf}
\end{table*}

\section{\label{sec:Injection}Sensitivity for Astrophysical Source Populations} 

In this section
we evaluate the performance of the search 
using simulated signals with the morphologies expected from core-collapse supernovae (CCSNe)~\citep{Janka2012,Gossan2016,PhysRevD.104.102002, SNGW_review} and sudden excitations of neutron stars~\citep{Watts2007,Giliberti2022}, of the type that might be associated with pulsar timing glitches and underlying starquakes~\footnote{This all-sky search also could, in principle, detect IMBH mergers or highly eccentric binary mergers, but we do not evaluate those signal classes in this paper, instead leaving them for future dedicated analyses.}.
This enables the interpretation of the null results of the present search in terms of astrophysically-motivated sources, most of which were also discussed in the previous LVK search on O3 data~\citep{O3-allsky-short}.
The procedure is based on the measurement of the detection efficiency: 
simulated signals are injected via software into the detector data streams and the analysis is repeated. Differently from the ad-hoc injections discussed in the previous section, here signal waveforms and amplitudes are informed by GW emission models and assumed spatial distributions of the sources.
This search is still all-sky, meaning that the search does not depend on having additional information on the source from other cosmic messengers, such as from electromagnetic (EM) or neutrino observations. If such information \emph{is} available to constrain the search, that would enable a deeper exploitation of the GW data.
The present all-sky search covers the scenario in which the EM emission is obscured or the EM or neutrino signal is intrinsically too weak to be detected.
Throughout this section, quoted sensitivities follow the criterion for significant detection of IFAR $\geq$ 100 years.

\subsection{\label{sec:CCSN}Core-Collapse Supernovae}

GW transients from CCSNe can show complex time-frequency structures related to the inner dynamics of the source, see e.g.\ \cite{SNGW_review}. 
After a \textit{bounce} which halts the initial collapse of the core, the continuing emission typically has a stochastic character with spectral features evolving in time.
The GW emission is modeled by numerical simulations of the hydrodynamics and microphysics of the source, which provide deeper comprehension of the rich physical processes occurring within the time scale of $\sim$1 second after the collapse of the progenitor star.
In general terms, the GW emission is dominated by the proto-neutron star, which is excited and deformed by many concurrent processes, including aspherical hydrodynamical flows developing within the central few hundred kilometers and rotation of the collapsing star. Most of these GW features fall inside the signal parameter space targeted by the low-frequency analyses used in the present search.

We evaluate the search's
sensitivity to CCSNe using predicted GW waveforms from seven different three-dimensional CCSN simulations. Along with the five models analyzed in the all-sky search on O3 data~\citep{O3-allsky-short}, we add two recent models of CCSNe 
which are likely to lack EM emission:
\begin{itemize}
 \item Model \textit{40 NR} (\textit{Pan+21 40 NR} from \cite{Pan_2021}) has a non-rotating, 40~$M_{\odot}$ zero age main sequence (ZAMS) mass progenitor with solar metallicity. Black hole (BH) formation occurs at $\sim$0.78~s after the initial bounce. The dominant GW emission is associated with the proto-neutron star excitation, shows a rising frequency with time, and peaks at $\sim$2~kHz before BH formation. A subdominant emission from standing accretion shock instability (SASI, \cite{SASI1, SASI2, SASI3}) is also present below 200~Hz.
 \item Model \textit{s18np} (\textit{Pow+20 s18np} from \cite{Powell_2020}) has a non--rotating, 18~$M_{\odot}$ ZAMS progenitor with solar metallicity. The GW emission from the proto-neutron star peaks at $\sim$0.7~kHz with a subdominant emission below 200~Hz from SASI. Differently from model  \textit{s18} below, there is no hint for shock revival by the end of the simulation (at  $\sim$0.56~s after the bounce).
\end{itemize}
The likely lack of EM emission makes these two models and model \textit{m20p} (see below) particularly interesting as plausible targets for an all-sky burst search.

The five models that were evaluated with the previous all-sky search (see \citep{O3-allsky-short} for more details) enable a direct comparison of sensitivity. 
They cover three cases with GW energy emission in the lower range of predictions and two cases of more extreme models with higher GW emission. The former three models have solar metallicity non-rotating progenitors:
\begin{itemize}
 \item Model \textit{s9} (\textit{Rad+19 s9} from \cite{Radice_2019}) from a 9~$M_{\odot}$ ZAMS progenitor. This is at the low end of the progenitor mass spectrum for CCSNe. 
 \item Model \textit{s18} (\textit{Pow+19 s18} from \cite{Powell_2018}) from a 18~$M_{\odot}$ ZAMS progenitor. Differently from model \textit{s18np}, this model includes strong aspherical seed perturbations from convective oxygen burning and undergoes shock revival and SN explosion.  
 \item Model  \textit{m20p} (\textit{Oco+18 mesa20\_3D\_pert} from  \cite{OConnor_Couch_2018}) from a 20~$M_{\odot}$ ZAMS progenitor. This model produces a BH remnant without powering EM emission. 

\end{itemize}
Extreme models with a strongly rotating, low-metallicity progenitor are:
\begin{itemize}
 \item Model \textit{m39}  (\textit{Pow+20 m39} from \cite{Powell_2020}) from a 39~$M_{\odot}$ ZAMS progenitor with 1/50 solar metallicity.
 \item Model \textit{35OC} (\textit{Obe+20 35OC-RO} from \cite{Obergaulinger_2020}) from a 35~$M_{\odot}$ ZAMS progenitor with 1/50 solar metallicity. 
\end{itemize}

\begin{figure}[t!]
  \begin{center}
  \includegraphics[width=1.0\columnwidth]{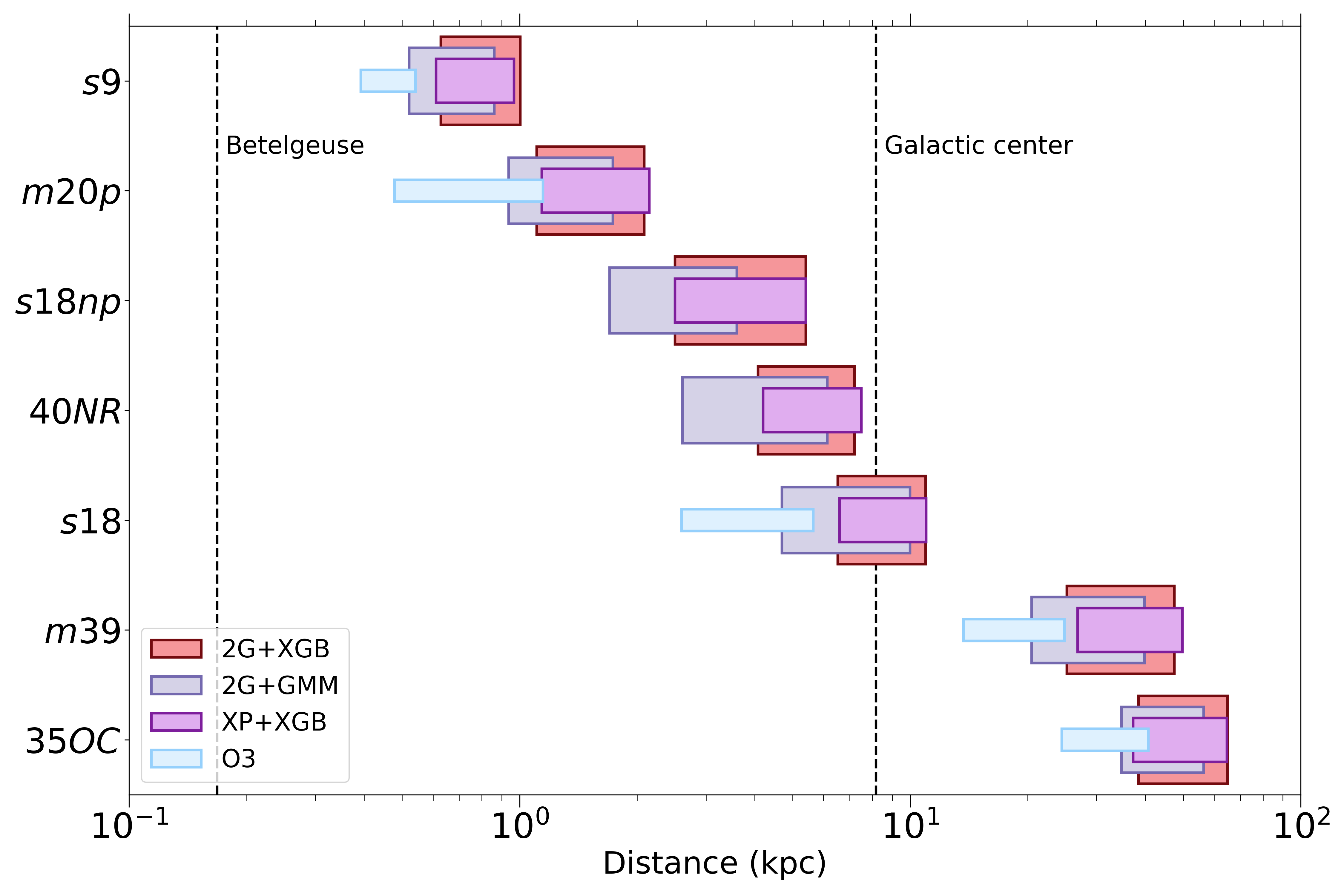}
  \end{center}
  \caption{Sensitivity of the all-sky GW search to core-collapse supernova events for different emission models. Boxes represent distance ranges between detection efficiencies 50\% and 10\% at IFAR $> 100$ years, evaluated using an isotropic distribution of sources, for each low-frequency analysis. We also plot the results from the previous observing run, O3. The current search reaches the Galactic center for three of the emission models we considered. Models \textit{40 NR} and \textit{m20p} form a BH remnant. The difference for models  \textit{s18} and \textit{s18np} reflects the impact of assuming diverse aspherical seed perturbations for progenitor stars with equal masses.
}
  \label{Fig.CCSN-Sensitivity}
\end{figure}

All signals from these models have been injected into the detectors' data streams with a random orientation of the source. The other extrinsic parameters of the source, in particular distance and sky direction, have been drawn from two population models: i) an isotropic model and ii) a Galactic model. The former allows to measure the average detection efficiency as a function of distance for each all-sky analysis. The sky direction is sampled uniformly in solid angle over all directions, while the distance to the source is distributed between 0.01 kpc and 100 kpc. We present in Figure~\ref{Fig.CCSN-Sensitivity} the distances corresponding to 50\% and 10\% detection efficiency. Our results improve the distance reach by a factor $\sim$2 for the five waveforms that are common to this work and the previously published O3 search \citep{O3-allsky-short}.

We note that
the LVK recently performed a dedicated search for GW bursts associated with SN 2023ixf~\cite{2023TNSTR1158} using LIGO data from the engineering run preceding the start of O4a. In that case, the published LVK results~\cite{LVK-SN2023ixf} make use of the known sky position of the source, achieving a somewhat better detection efficiency at equal distance for the four waveform models in common. Compared to the ad-hoc signals discussed in section V,  CCSNe signals are more challenging to detect due to their wider time-frequency structure, as  discussed in~\cite{PhysRevD.104.102002, PhysRevD.106.063014}.

In the Galactic model, we generate a population of simulated sources with sky directions and distances drawn from a model of the Milky Way stellar mass distribution, including a bulge, a thick stellar disk and a thin stellar disk, whose parameters are taken from \cite{MilkyWay1} and \cite{MilkyWay2}. In this case we present the overall efficiencies of the three low-frequency analyses for each of the CCSN models in Table~\ref{Tab.O4aGalSN}. The efficiency is representative of the probability of detection of a Galactic CCSN for each model: we call this CCSN Galactic coverage.
For two of the lower-energy CCSN models (\textit{s9} and \textit{m20p}) the Galactic coverage is practically zero. In fact, \textit{s9} has a progenitor at the lower end of the mass spectrum of CCSNe and \textit{m20p} lacks shock revival, which make them detectable only out to $\sim$1~kpc. Within this range, the Galactic distribution provides very rare CCSNe.
On the other hand, the Galactic coverage is $\sim$90\% or higher for the \textit{m39} and \textit{35OC} models, meaning that a CCSN with those properties occuring at a random location in the Milky Way during O4a would likely have been detected by our search.

\begin{table}[!htb]
\begin{center}
\begin{tabular}{c | c c c | c}
\hline
\hline
\textbf{Model} & \ac{2G} & \ac{GMM} & \ac{XP} & best O3 \\
\hline
\textit{s9} & $< 0.1\%$ & $< 0.1\%$ & $< 0.1\%$ & $< 0.1\%$\\
\textit{m20p} & $< 0.1\%$ & $< 0.1\%$ & $< 0.1\%$ & $< 0.1\%$\\
\textit{s18np} & $4.6\%$ & $2.0\%$ & $4.5\%$ & --\\
\textit{40 NR} & $20.4\%$ & $19.6\%$ & $22.8\%$ & --\\
\textit{s18} & 34.1\% & 20.4\% & 32.3\% & 1.2\%\\
\textit{m39} & 89.5\% & 74.0\% & 87.4\% & 69.4\%\\
\textit{35OC} & 97.2\% & 93.8\% & 95.7\% & 89.8\%\\
\hline
\hline
\end{tabular}
\end{center}
\caption{The fraction of Galactic CCSN signals that are detectable with an IFAR threshold of 100 years or greater by each low-frequency analysis, for different modeled CCSN waveforms. There is a huge difference between models of CCSNe with lower and higher GW amplitudes.
The last column reports the best results on O3 data~\citep{O3-allsky-short}.} \label{Tab.O4aGalSN}
\end{table}

\subsection{\label{sec:NSglitch} Isolated Neutron Star Emitters}

Among the known pulsars (rotating neutron stars with beamed radio and/or high-energy EM emission that is detected at Earth as pulses at nearly constant intervals), some exhibit sudden increases in pulse rate known as timing \emph{glitches}.
Such a change indicates a reconfiguration of the compact object and its magnetosphere with a concurrent release of energy, some of which may go into transient GWs.
In this work we consider the detectability of GW transients connected with neutron star timing glitches from the inferred population of nearby neutron stars which are not observed as pulsars because their EM emission is not angled favorably for telescopes on Earth to detect.
The GW emission from a perturbed, rotating neutron star is more isotropic and can be detected even from EM-dark pulsars \cite{Lopez:2022yph}. 
Transient excitations of neutron stars associated with \emph{observed} EM signatures are best covered by dedicated GW searches, e.g.\ for the August 2006 glitch of the Vela Pulsar \cite{S6_NS} or magnetar flares (see \cite{magnetarsO3_Abbott_2024} for O3 results).

To investigate the sensitivity of our search to neutron star excitations, we follow
the same lines adopted in the previous LVK analysis on O3 data~\citep{O3-allsky-short}. In particular, we assume that the dominant GW emission is due to a short-duration \textit{f}-mode excitation of the glitching neutron star \cite{S6_NS,WH} and we model the related GW burst as a damped sinusoid signal. This choice is not restrictive, however, as our search does not depend on the morphology of the incoming GW signal at first order. Typical frequencies and damping times of these signals are expected to be in the ranges  $2 \ \mathrm{kHz} \leq \nu_{\mathrm{GW}} \leq 3 \ \mathrm{kHz}$ and from tens of milliseconds to as much as half a second respectively, depending on the mass and the equation of state (EoS) of the neutron star \cite{WH}. Searches for GW emission with much longer duration have been performed using different methods, see e.g.\ the LVK results on O3 data \cite{Abbott_2022_PSR_narrowbandCW}.

Our all-sky search has not found any significant event in this high-frequency band. The null result can be interpreted under the optimistic asumptions that all the glitch energy is converted into GW energy,
$E_{\mathrm{glitch}}=E_{\mathrm{GW}}$, and that the source's angular momentum is optimally oriented, so that the incoming GW burst is circularly polarized. 
We adopt as a standard source a pulsar at a distance of 287 pc and with spin frequency $\nu_s = 11.2$ $\text{Hz}$, which are the parameters of the closest known glitching pulsar~\cite{vdist, vspin}, the Vela Pulsar.
This allows us to interpret the GW search in terms of minimum detectable glitch size,  $\Delta \nu_{s}$, following equation 5 in \cite{WH}.
The Vela Pulsar belongs to the subpopulation of known pulsars which are both spinning rapidly and undergoing larger glitches~\cite{vdist, vspin}, and thus are delivering glitch energies at the higher end of the observed range, since $E_{\mathrm{glitch}} \propto \nu_s \Delta \nu_{s}$. 

\begin{figure}[t!]
  \begin{center}
  \includegraphics[width=1.0\columnwidth]{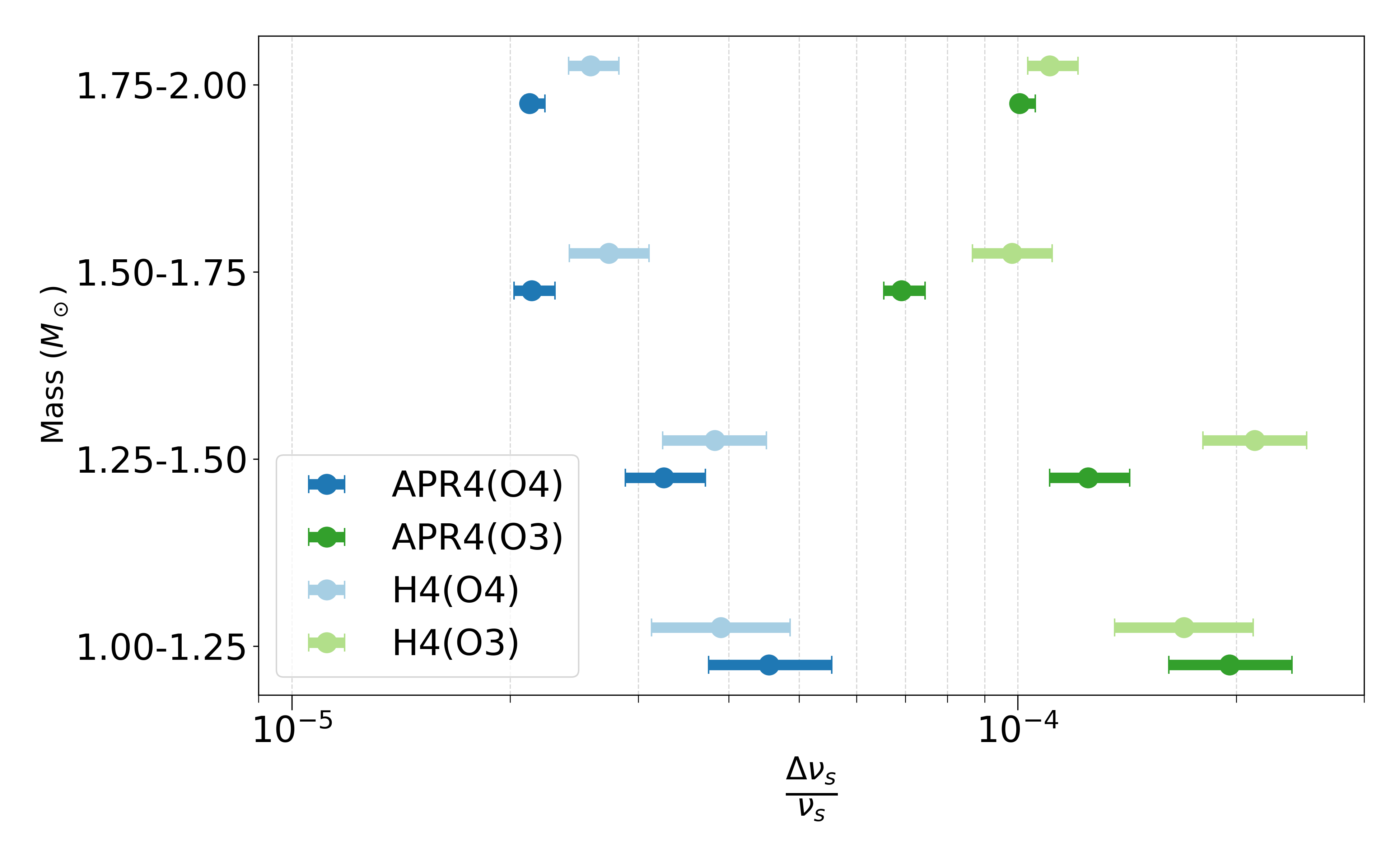}
  \end{center}
  \caption{ Sensitivity of the high-frequency \ac{2G} analysis to neutron star glitches, considering spin frequency and distance of the Vela Pulsar as a representative standard source for soft (APR4) and hard (H4) EoS and assuming an optimally oriented source at random sky directions.  
  For each EoS and mass bin, the dot shows the mean value of the glitch size that yields 50\% detection efficiency at IFAR $\geq$ 100 years, and the error bar represents the variation within the mass bin. On average, a higher-mass neutron star allows for smaller glitches to be detected. The fractional glitch size $\Delta \nu_{s}/\nu_s$ for a Vela-like pulsar would need to be stronger than $ \sim 2\,\text{to}\,6\,\times 10^{-5}$ for 50\% of the sources to be detected in O4a.}
  \label{Fig.NS_sensitivity}
\end{figure}

A population of such standard sources is simulated uniformly distributed in the sky and emitted GW signals are estimated according to two EoS models, APR4 (soft) \cite{apr4} and H4 (hard) \cite{h4}, for neutron star masses  in the range of 1--2 $M_{\odot}$ in the non-rotating limit \cite{MR-FT}.
Detectable glitch sizes (requiring IFAR $\geq$ 100 years) are reported as a function of the neutron star mass and EoS in Figure~\ref{Fig.NS_sensitivity}.
The minimum detectable pulsar glitch size for the O4a run, expressed as a fractional change referenced to the Vela Pulsar's frequency, $\Delta \nu_{s}/\nu_s$, ranges from $\sim 2 \times 10^{-5}$ to $\sim 6 \times 10^{-5}$ depending on the neutron star mass. These values are a factor of $\sim$3 to 5 better than what was obtained in O3. The actual Vela Pulsar glitches with fractional magnitude $\Delta \nu_{s}/\nu_s \approx 1 \textup{--} 3 \times 10^{-6}$ approximately every 3 years \cite{vela_p}, so  this all-sky search is still an order of magnitude above that value. 
Nevertheless, many known pulsars show glitches in the range $10^{-5} \textup{--} 10^{-4}\,\text{Hz}$ \cite{gref_1,gref_2,gref_3}, therefore our current sensitivity provides a potential discovery channel for yet unknown glitching pulsars which could be closer to Earth and/or spinning faster than the Vela Pulsar.
\label{ssec:NS_sensitivity}

\section{\label{sec:Conclusions}Conclusions}

We reported the results of a search for gravitational-wave bursts in LIGO data during the LVK O4a observing period using three pipelines based on variants of the cWB algorithm followed by machine-learning stages. 
Three low-frequency analyses and one high-frequency analysis targeted GW bursts lasting up to a few seconds with frequency content in the range $16$--$4096$ Hz.
By searching as broadly as possible, we sought to uncover new types of gravitational-wave signals from either known or unknown astrophysical objects.
No significant candidate was found beyond the CBC candidates already identified by low-latency searches using modeled templates.
The most significant new candidate found has a false alarm rate of one per 1.75 years,
statistically consistent with the expected background for this dataset.

We evaluated the sensitivity of the search using ad-hoc simulated signals as well as two types of astrophysically motivated signals.
The ad-hoc signals consisted of sine-Gaussians with a wide range of frequencies and $Q$ values plus a few Gaussian pulses and white-noise bursts.
Our search was capable of detecting such signals at amplitudes typically a factor of 2 or more weaker than the previous published search, which used data from the LVK O3 run.
Thus, even with the shorter duration of O4a, our search achieved the greatest time-volume exposure to date for generic short-duration burst signals.
Assuming a population of rotating sources, we placed rate density upper limits an order of magnitude lower than the O3 search for signal frequencies between $\sim$100 Hz and 1 kHz.
We considered seven stellar core-collapse models and found that our search could have detected GW signals throughout most of the Milky Way from two of the models with high-mass star progenitors,
with more limited reach for the other models.
Finally, we simulated damped sinusoid GW signals representing neutron star \textit{f}-mode oscillations excited in connection with pulsar timing glitches.
Using an optimistic energy argument, we estimated the glitch size (fractional frequency change) that our GW search could have detected, finding that typical glitches from the Vela Pulsar would not have been detectable but a similarly close unseen pulsar with plausible properties could potentially have been detected if it had experienced a timing glitch during O4a.

This search only used data from the first period of the O4 run, from May 24, 2023 to January 16, 2024. After a commissioning break, the O4 run resumed on April 10, 2024 with the Virgo detector collecting data along with the two LIGO detectors.
While Virgo has been a factor of $\sim$3 less sensitive than LIGO during the O4 run, it often helps greatly to localize event candidates in the sky by triangulation with the three-detector network \cite{ObservingScenarios2020,Lopez:2022yph}.
The KAGRA detector rejoined the O4 run in June 2025 with a noise level $\sim$7 times less sensitive than Virgo, as commissioning of KAGRA is still ongoing.
The O4 run is now expected to end on November 18, 2025 after accumulating about two years of GW data, excluding commissioning breaks during the run.
The full O4 data set will enable more-sensitive searches for GW burst signals,
and having data from three (or more) sensitive detectors will improve our ability to characterize any candidate which is found.
Even if O4 does not reveal any GW burst signals,
subsequent observing runs after further detector upgrades, and with new facilities, will enable deeper, longer-duration searches.

\begin{acknowledgments}
This material is based upon work supported by NSF's LIGO Laboratory, which is a
major facility fully funded by the National Science Foundation.
The authors also gratefully acknowledge the support of
the Science and Technology Facilities Council (STFC) of the
United Kingdom, the Max-Planck-Society (MPS), and the State of
Niedersachsen/Germany for support of the construction of Advanced LIGO 
and construction and operation of the GEO\,600 detector. 
Additional support for Advanced LIGO was provided by the Australian Research Council.
The authors gratefully acknowledge the Italian Istituto Nazionale di Fisica Nucleare (INFN),  
the French Centre National de la Recherche Scientifique (CNRS) and
the Netherlands Organization for Scientific Research (NWO)
for the construction and operation of the Virgo detector
and the creation and support  of the EGO consortium. 
The authors also gratefully acknowledge research support from these agencies as well as by 
the Council of Scientific and Industrial Research of India, 
the Department of Science and Technology, India,
the Science \& Engineering Research Board (SERB), India,
the Ministry of Human Resource Development, India,
the Spanish Agencia Estatal de Investigaci\'on (AEI),
the Spanish Ministerio de Ciencia, Innovaci\'on y Universidades,
the European Union NextGenerationEU/PRTR (PRTR-C17.I1),
the ICSC - CentroNazionale di Ricerca in High Performance Computing, Big Data
and Quantum Computing, funded by the European Union NextGenerationEU,
the Comunitat Auton\`oma de les Illes Balears through the Conselleria d'Educaci\'o i Universitats,
the Conselleria d'Innovaci\'o, Universitats, Ci\`encia i Societat Digital de la Generalitat Valenciana and
the CERCA Programme Generalitat de Catalunya, Spain,
the Polish National Agency for Academic Exchange,
the National Science Centre of Poland and the European Union -- European Regional
Development Fund;
the Foundation for Polish Science (FNP),
the Polish Ministry of Science and Higher Education,
the Swiss National Science Foundation (SNSF),
the Russian Science Foundation,
the European Commission,
the European Social Funds (ESF),
the European Regional Development Funds (ERDF),
the Royal Society, 
the Scottish Funding Council, 
the Scottish Universities Physics Alliance, 
the Hungarian Scientific Research Fund (OTKA),
the French Lyon Institute of Origins (LIO),
the Belgian Fonds de la Recherche Scientifique (FRS-FNRS), 
Actions de Recherche Concert\'ees (ARC) and
Fonds Wetenschappelijk Onderzoek -- Vlaanderen (FWO), Belgium,
the Paris \^{I}le-de-France Region, 
the National Research, Development and Innovation Office of Hungary (NKFIH), 
the National Research Foundation of Korea,
the Natural Sciences and Engineering Research Council of Canada (NSERC),
the Canadian Foundation for Innovation (CFI),
the Brazilian Ministry of Science, Technology, and Innovations,
the International Center for Theoretical Physics South American Institute for Fundamental Research (ICTP-SAIFR), 
the Research Grants Council of Hong Kong,
the National Natural Science Foundation of China (NSFC),
the Israel Science Foundation (ISF),
the US-Israel Binational Science Fund (BSF),
the Leverhulme Trust, 
the Research Corporation,
the National Science and Technology Council (NSTC), Taiwan,
the United States Department of Energy,
and
the Kavli Foundation.
The authors gratefully acknowledge the support of the NSF, STFC, INFN and CNRS for provision of computational resources.

This work was supported by MEXT,
the JSPS Leading-edge Research Infrastructure Program,
JSPS Grant-in-Aid for Specially Promoted Research 26000005,
JSPS Grant-in-Aid for Scientific Research on Innovative Areas 2402: 24103006,
24103005, and 2905: JP17H06358, JP17H06361 and JP17H06364,
JSPS Core-to-Core Program A.\ Advanced Research Networks,
JSPS Grants-in-Aid for Scientific Research (S) 17H06133 and 20H05639,
JSPS Grant-in-Aid for Transformative Research Areas (A) 20A203: JP20H05854,
the joint research program of the Institute for Cosmic Ray Research,
University of Tokyo,
the National Research Foundation (NRF),
the Computing Infrastructure Project of the Global Science experimental Data hub
Center (GSDC) at KISTI,
the Korea Astronomy and Space Science Institute (KASI),
the Ministry of Science and ICT (MSIT) in Korea,
Academia Sinica (AS),
the AS Grid Center (ASGC) and the National Science and Technology Council (NSTC)
in Taiwan under grants including the Science Vanguard Research Program,
the Advanced Technology Center (ATC) of NAOJ,
and the Mechanical Engineering Center of KEK.

Additional acknowledgements for support of individual authors may be found in the following document: \\
\texttt{https://dcc.ligo.org/LIGO-M2300033/public}.
For the purpose of open access, the authors have applied a Creative Commons Attribution (CC BY)
license to any Author Accepted Manuscript version arising.
We request that citations to this article use ``A.\ G.\ Abac {\it et al.}\ (LIGO-Virgo-KAGRA Collaboration), ...'' or similar phrasing, depending on journal convention.

This paper has been assigned LIGO document ID P2400601-v8.
\end{acknowledgments}

\appendix

\bibliography{o4ashort}

\end{document}